# Impact of Security Modeling Fidelity on Hyperscale Data Center Hosting Capacity Assessment

Haoxiang Wan[1], Linhan Fang[1], Jonghwan Kwon[2], Jesus Silva-Rodriguez[3], Jianyi Yang[4], Jaya Yellajosula[5], Fan Jiang[1], and Xingpeng Li[1]

1 Department of Electrical and Computer Engineering, University of Houston, Houston, TX, USA
2 Energy Systems and Infrastructure Analysis Division, Argonne National Laboratory, Lemont, IL, USA
3 Electric Reliability Council of Texas, Taylor, TX, USA
4 Department of Computer Science, University of Houston, Houston, TX, USA
5 Sargent and Lundy, Houston, TX, USA

***Abstract*— This paper assesses hyperscale data center hosting capacity in transmission grids. Hosting capacity (HC) analysis quantifies the maximum additional load a bus can accommodate without violating operational constraints. This paper defines data center HC as the maximum time-invariant active power demand at a single target bus over a 24-hour peak-day horizon, and evaluates it under four models of increasing fidelity (i.e., the extent to which the model enforces real-world operational security requirements) on the modified IEEE 24-bus system: (i) DC optimal power flow (DC-OPF) without unit commitment (UC), (ii) basic SCUC (security-constrained unit commitment) without reliability constraints, (iii) SCUC with spinning reserve, and (iv) *N*-SCUC with both reserve and branch-contingency security. Comparing these models quantifies the impact of unit commitment, reserve requirements, and *N*-1 security on bus-level HC estimates. The *N*-SCUC yields an average HC of 444 MW; omitting *N*-1 security raises this to 532 MW, a 19.8% overestimation. The DC-OPF and basic SCUC models further overestimate HC, indicating a trade-off between model complexity and estimation accuracy.**

## NOMENCLATURE

*Sets and Indices*

| | |
|---|---|
| $\mathcal{N}$ | Set of buses, indexed by $n$. |
| $\mathcal{L}$ | Set of branches (transmission lines), indexed by $\ell$. |
| $\mathcal{G}$ | Set of generators, indexed by $g$ (also $s$). |
| $\mathcal{T}$ | Set of hourly time periods, $\{1, \ldots, T\}$, $T = 24$, indexed by $t$. |
| $\mathcal{C}$ | Set of *N*-1 branch contingencies (non-radial branches), indexed by $c$. |
| $\mathcal{G}_n$ | Subset of generators located at bus $n$, $\mathcal{G}_n \subseteq \mathcal{G}$. |
| $\mathcal{L}_n^{\text{to}}$ | Branches whose receiving end is bus $n$. |
| $\mathcal{L}_n^{\text{from}}$ | Branches whose sending end is bus $n$. |

*Parameters*

| | |
|---|---|
| $P_g^{\max}$ | Maximum generation output of unit $g$. |
| $P_g^{\min}$ | Minimum generation output of unit $g$. |
| $HR_g$ | Hourly ramp rate of unit $g$. |
| $SR_g$ | Startup ramp rate of unit $g$. |
| $SPR_g$ | 10-minute spinning ramp capability of unit $g$. |
| $MU_g$ | Minimum up time of unit $g$. |
| $MD_g$ | Minimum down time of unit $g$. |
| $B_\ell$ | Susceptance of branch $\ell$. |
| $n(\ell)$ | From-bus (sending-end) index of branch $\ell$. |
| $m(\ell)$ | To-bus (receiving-end) index of branch $\ell$. |
| $F_\ell^A$ | Rate A (normal) thermal limit of branch $\ell$. |
| $F_\ell^C$ | Rate C (emergency) thermal limit of branch $\ell$. |
| $D_{n,t}$ | Demand at bus $n$ in hour $t$; $D_{n,t} = D_n^{\text{base}} \cdot \lambda_t/100$. |
| $D_n^{\text{base}}$ | Base (peak) load at bus $n$. |
| $\lambda_t$ | Hourly load multiplier at time $t$. |
| $\lambda_{\max}$, $\lambda_{\min}$ | Peak and minimum load multiplier values. |
| $A_{c,\ell}$ | Contingency-branch indicator; $A_{c,\ell} = 0$ if $c = \ell$ (outaged), else 1. |
| $T$ | Total number of hourly periods in the scheduling horizon. |
| $b^*$ | Target bus at which hosting capacity is maximized. |
| $\mathbb{1}_{[\cdot]}$ | Indicator function; equals 1 if the condition holds, 0 otherwise. |

*Decision Variables*

| | |
|---|---|
| $P^{\text{HC}}$ | Hosting capacity at target bus $b^*$, time-invariant across all hours. |
| $P_{g,t}$ | Active power output of generator $g$ at hour $t$. |
| $\theta_{n,t}$ | Voltage angle at bus $n$ at hour $t$. |
| $P_{\ell,t}^f$ | Active power flow on branch $\ell$ at hour $t$. |
| $u_{g,t}$ | Binary commitment status of unit $g$ at hour $t$; $\in \{0,1\}$. |
| $v_{g,t}$ | Binary startup indicator of unit $g$ at hour $t$; $\in \{0,1\}$. |
| $w_{g,t}$ | Binary shutdown indicator of unit $g$ at hour $t$; $\in \{0,1\}$. |
| $R_{g,t}$ | Spinning reserve provided by unit $g$ at hour $t$. |
| $P_{g,c,t}^c$ | Post-contingency output of unit $g$, contingency $c$, hour $t$. |
| $P_{\ell,c,t}^{f,c}$ | Post-contingency flow on line $\ell$, contingency $c$, hour $t$. |
| $\theta_{n,c,t}^c$ | Post-contingency angle at bus $n$, contingency $c$, hour $t$. |

## I. INTRODUCTION

Hosting capacity (HC) has become a key planning metric for utilities seeking to evaluate the ability of their networks to accommodate additional load or distributed energy resources without requiring infrastructure upgrades [1]. HC is typically defined as the maximum incremental active power injection or withdrawal at a given bus that can be sustained while satisfying all operational constraints, including thermal line limits, voltage limits, and generation adequacy [2]. Large data center loads, in particular, are expanding rapidly. North American Electric Reliability Corporation (NERC) identifies data center growth as a primary driver of an unprecedented 132 GW (15%) ten-year summer peak demand increase in North America [3]. A complementary bottom-up analysis by Lawrence Berkeley National Laboratory estimates that U.S. data center electricity consumption will rise from 176 TWh in 2023 (4.4% of national electricity use) to 325–580 TWh by 2028 [4]. Thus, accurate

HC assessment at the transmission level is increasingly important for infrastructure investment planning and interconnection studies [5], [6].

The security-constrained unit commitment (SCUC) problem is a cornerstone of power system operations, jointly determining generator scheduling and dispatch while enforcing *N*-1 contingency security and reserve requirements [7]. Recent extensions further incorporate locational frequency stability for low-inertia grids [8], and considerable effort has focused on reducing the computational burden of *N*-SCUC through decomposition [9] and machine learning-based screening [10]. Evaluating HC within an SCUC framework yields operationally realistic estimates, but at substantial cost: a single *N*-SCUC instance involves hundreds of post-contingency scenarios, which becomes prohibitive when many candidate buses must be screened [10]. In current industry practice, utilities and RTOs/ISOs assess generator interconnection capability primarily through AC power-flow contingency studies, which screen thermal and voltage violations under planning reliability criteria [11], [12]. FERC Order No. 2023 further requires transmission providers to post heatmaps of estimated incremental injection capacity at each bus under *N*-1 conditions [13]. The proposed SCUC framework is complementary: it captures chronological commitment, dispatch, reserves, and *N*-1 active-power security, but its DC network model does not represent reactive power or voltage magnitudes. Thus, the reported HC should be interpreted as commitment- and dispatch-feasible capacity, subject to final AC contingency validation.

The resulting HC values depend heavily on which security layers are enforced—*N*-1, reserve, and full unit commitment (UC) logic at the high-fidelity end, versus continuously available generation and no post-contingency checks at the simplified end. More restrictive constraints reduce available transfer capability and thus HC. Yet few studies systematically quantify how each layer contributes to bus-level HC; most adopt a single fixed formulation or target generation interconnection rather than load hosting [11].

This paper focuses on a controlled comparison of how progressive security-modeling layers affect bus-level HC under a common HC definition and test system. Specifically, four optimization models are formulated on the modified IEEE Reliability Test System (RTS) 24-bus system and solved independently for each target bus. For each run, one of the 24 buses is designated as the target bus, the additional load at that bus is maximized, and all other buses retain their nominal demand, yielding a complete spatial map of HC. The four models, ordered by increasing fidelity, are:

- **DC-OPF**: DC optimal power flow, a simplified linear model without UC, reserve, or *N*-1 constraints.
- **B-SCUC**: basic SCUC with neither reserve requirement nor *N*-1 constraints.
- **R-SCUC**: regular SCUC with reserve requirement but without *N*-1 contingency constraints.
- ***N*-SCUC**: enhanced SCUC with both spinning reserve requirement and *N*-1 reliability constraints [14].

The primary contributions of this paper are as follows:

- A systematic comparison framework is developed that applies four formulations of increasing fidelity, controlled by the inclusion of *N*-1 branch contingencies, spinning reserve, and UC constraints, to a common HC definition and test system, isolating the individual effect of each security layer on bus-level HC.
- HC values are computed at the nodal level with each model, enabling a comprehensive comparison of how the HC spatial distribution across buses changes with model fidelity.
- The accuracy cost of each simplification is quantified, providing quantitative benchmarks that let planners use a tractable simplified model as an optimistic HC upper bound when the *N*-SCUC is intractable.

The remainder of this paper is organized as follows. Section II describes the system model, the HC definition adopted, and the complete mathematical formulation of the four models. Section III reports and discusses the results. Section IV concludes the paper and outlines limitations and future work.

## II. System Model and Formulation

### A. HC Definition and Test System

The hosting capacity $P^{\mathrm{HC}}$ is modeled as a single scalar decision variable representing an additional active power load at one target bus $b^*$, held constant across all $T = 24$ hours. This reflects a planning-stage convention in which the new load is represented by its sustained (firm) component [13]; $P^{\mathrm{HC}}$ must be feasible in every hour and is therefore upper bounded by the most stressed hour, typically system peak hours. For each formulation (DC-OPF, B-SCUC, R-SCUC, and *N*-SCUC), the optimization is solved independently for each possible data center installation location among the 24 buses; only $b^*$ supplies the additional data center load while all other buses retain nominal demand. Reported per-bus values therefore characterize each bus in isolation and should not be summed across buses.

The modified IEEE RTS 24-bus system is adopted as the test network, comprising 24 buses, 33 generators, and 38 branches (33 transmission lines and 5 transformers), spanning a 138-kV sub-network (Buses 1–10) and a 230-kV sub-network (Buses 11–24). Total installed generation is 3,405 MW and total base load is 1,887 MW, with seven buses carrying zero base load. A 24-hour horizon is modeled using hourly load multipliers $\lambda_t$ that scale the base demand: $\lambda_{\max} = 121\%$ for system peak hours of approximately 2,283 MW and $\lambda_{\min} = 65\%$ for grid off-peak hours of approximately 1,227 MW. The DC power flow approximation is employed throughout. The *N*-1 set $\mathcal{C}$ in *N*-SCUC is restricted to *branch* contingencies; Branch 11 (Buses 7–8) is radial and excluded from $\mathcal{C}$, leaving $|\mathcal{C}| = 37$ scenarios. Generator *N*-1 and multi-element *N*-2 contingencies are deferred to future work. The four models are summarized in Table I and ordered by increasing fidelity.

### B. Mathematical Formulation

All quantities in the formulation below are expressed in MW and numerical results in Section III are also reported in MW.

Table I. Summary of the Four Models and Their Problem Size on the Modified IEEE RTS 24-Bus System

| Feature | DC-OPF* | B-SCUC | R-SCUC | *N*-SCUC |
|---|---|---|---|---|
| Objective (1) | ✓ | ✓ | ✓ | ✓ |
| Generator & UC (2)–(4)*, (8)–(17) | (5)* | ✓ | ✓ | ✓ |
| Spinning Reserve (6)–(7) | – | – | ✓ | ✓ |
| *N*-1 Contingency (22)–(27) | – | – | – | ✓ |
| Variables | 2,281 | 4,657 | 5,449 | 89,809 |
| Constraints | 4,920 | 9,311 | 10,895 | 250,655 |
| Binary Variables | 0 | 2,376 | 2,376 | 2,376 |

*In DC-OPF, the unit-commitment output bounds (2)–(4) are replaced by continuous generation bounds (5), since no units are committed; the network constraints (18)–(21) apply to all four models.

*Objective function:*

$$\max\ P^{\mathrm{HC}} \tag{1}$$

*Generator and unit commitment constraints:*

$$P_g^{min} \cdot u_{g,t} \le P_{g,t}, \quad \forall g,\ t \tag{2}$$

$$P_{g,t} + R_{g,t} \le P_g^{max} \cdot u_{g,t}, \quad \forall g,\ t \tag{3}$$

$$P_{g,t} \le P_g^{\mathrm{max}} \cdot u_{g,t}, \quad \forall g,\ t \tag{4}$$

*Continuous generation bounds (DC-OPF only, replacing (2)–(4)):*

$$P_g^{\mathrm{min}} \le P_{g,t} \le P_g^{max}, \quad \forall g,\ t \tag{5}$$

$$R_{g,t} \le \mathrm{SPR}_g \cdot u_{g,t}, \quad \forall g,\ t \tag{6}$$

$$\sum_{s\in\mathcal{G}} R_{s,t} \ge P_{g,t} + R_{g,t}, \quad \forall g,\ t \tag{7}$$

$$P_{g,t} - P_{g,t-1} \le \mathrm{HR}_g \cdot u_{g,t-1} + \mathrm{SR}_g \cdot v_{g,t}, \quad \forall g,\ t \ge 2 \tag{8}$$

$$P_{g,t-1} - P_{g,t} \le HR_g \cdot u_{g,t} + SR_g \cdot w_{g,t}, \quad \forall g,\ t \ge 2 \tag{9}$$

$$P_{g,1} - P_{g,T} \le \mathrm{HR}_g \cdot u_{g,T} + SR_g \cdot v_{g,1}, \quad \forall g \tag{10}$$

$$P_{g,T} - P_{g,1} \le HR_g \cdot u_{g,1} + SR_g \cdot w_{g,1}, \quad \forall g \tag{11}$$

$$\sum_{s=t-\mathrm{MU}_g+1}^{t} v_{g,s} \le u_{g,t}, \quad \forall g,\ t \ge MU_g \tag{12}$$

$$\sum_{s=T+t-\mathrm{MU}_g+1}^{T} v_{g,s} + \sum_{s=1}^{t} v_{g,s} \le u_{g,t}, \quad \forall g,\ t < MU_g \tag{13}$$

$$\sum_{s=t+1}^{t+\mathrm{MD}_g} v_{g,s} \le 1 - u_{g,t}, \quad \forall g,\ t \le T - MD_g \tag{14}$$

$$\sum_{s=1}^{t+\mathrm{MD}_g-T} v_{g,s} + \sum_{s=t+1}^{T} v_{g,s} \le 1 - u_{g,t}, \quad \forall g,\ t > T - MD_g \tag{15}$$

$$v_{g,t} - w_{g,t} = u_{g,t} - u_{g,t-1}, v_{g,t} + w_{g,t} \le 1, \quad \forall g,\ t \ge 2 \tag{16}$$

$$v_{g,1} - w_{g,1} = u_{g,1} - u_{g,T}, v_{g,1} + w_{g,1} \le 1, \quad \forall g \tag{17}$$

*Network constraints:*

$$\theta_{1,t} = 0, \quad \forall t \tag{18}$$

$$P_{\ell,t}^f = B_\ell \cdot \left(\theta_{n(\ell),t} - \theta_{m(\ell),t}\right), \quad \forall \ell,\ t \tag{19}$$

$$-F_\ell^A \le P_{\ell,t}^f \le F_\ell^A, \quad \forall \ell,\ t \tag{20}$$

$$\sum_{g\in\mathcal{G}_n} P_{g,t} + \sum_{\ell\in\mathcal{L}_n^{\mathrm{to}}} P_{\ell,t}^f - \sum_{\ell\in\mathcal{L}_n^{\mathrm{from}}} P_{\ell,t}^f = D_{n,t} + P^{\mathrm{HC}} \cdot \mathbb{1}_{[n=b^*]}, \quad \forall n,\ t \tag{21}$$

*N-1 security constraints (N-SCUC only):*

$$P_{g,t} - P_{g,c,t}^c \le \mathrm{SPR}_g \cdot u_{g,t}, \quad \forall g,\ c,\ t \tag{22}$$

$$P_{g,c,t}^c - P_{g,t} \le \mathrm{SPR}_g \cdot u_{g,t}, \quad \forall g,\ c,\ t \tag{23}$$

$$P_g^{\mathrm{min}} \cdot u_{g,t} \le P_{g,c,t}^c \le P_g^{\mathrm{max}} \cdot u_{g,t}, \quad \forall g,\ c,\ t \tag{24}$$

$$P_{\ell,c,t}^{f,c} = A_{c,\ell} \cdot B_\ell \cdot \left(\theta_{n(\ell),c,t}^c - \theta_{m(\ell),c,t}^c\right), \quad \forall \ell,\ c,\ t \tag{25}$$

$$-F_\ell^C \cdot A_{c,\ell} \le P_{\ell,c,t}^{f,c} \le F_\ell^C \cdot A_{c,\ell}, \quad \forall \ell,\ c,\ t \tag{26}$$

$$\sum_{g\in\mathcal{G}_n} P_{g,c,t}^c + \sum_{\ell\in\mathcal{L}_n^{\mathrm{to}}} P_{\ell,c,t}^{f,c} - \sum_{\ell\in\mathcal{L}_n^{\mathrm{from}}} P_{\ell,c,t}^{f,c} = D_{n,t} + P^{\mathrm{HC}} \cdot \mathbb{1}_{[n=b^*]}, \quad \forall n,\ c,\ t \tag{27}$$

The objective (1) maximizes the time-invariant large load $P^{\mathrm{HC}}$ at the target bus $b^*$. System operating costs (marginal cost, no-load cost, startup cost, and load-shedding penalties) are intentionally excluded from (1), because the purpose of this study is to determine the maximum feasible additional load each bus can physically host, not the cost-optimal dispatch. A cost-minimizing formulation would address a different question and would generally yield a lower HC, since expensive but feasible operating points would be avoided; such a cost-aware extension is left to future work. Generator output bounds depend on the model: when UC is active (B-SCUC, R-SCUC, and *N*-SCUC), constraint (2) enforces the minimum output through the binary commitment status $u_{g,t}$, while the upper bound is given by (3) with reserve (R-SCUC and *N*-SCUC); in B-SCUC, $R_{g,t}$ is fixed to zero, so (3) reduces to (4). In the DC-OPF model, all units are continuously available and (5) replaces the binary-dependent bounds. Constraints (6)–(7) define the unit-level and system-wide spinning reserve requirements (R-SCUC and *N*-SCUC); (7) is written in envelope form so that, for every $g$, the total system reserve covers the simultaneous loss of $g$'s output *and* its committed reserve, ensuring sufficient capacity to compensate for the largest committed unit. Hourly ramp-up and ramp-down limits are enforced by (8)–(9), with wrap-around constraints (10)–(11) enforcing cyclic consistency between the last and first hours. Minimum up- and down-time logic is captured by (12)–(15), again with wrap-around versions for the boundary hours, and the startup indicator $v_{g,t}$ is linked to commitment status by (16)–(17).

For the network, Bus 1 is the angle reference (18); the linearized DC power flow (19); base-case thermal limits are enforced by (20), and nodal power balance by (21), where $P^{\mathrm{HC}}$ enters only at the target bus $b^*$ and is identical across all hours by construction.

The *N*-1 security constraints (22)–(27) apply only to *N*-SCUC. The post-contingency angle reference is identical to the base case, with Bus 1 fixed at zero under all contingencies; in the implementation, this is imposed as $\theta_{1,c,t}^c = 0$, $\forall c \in \mathcal{C}$, $t \in \mathcal{T}$. Post-contingency re-dispatch is bounded by the 10-minute spinning ramp capability (22)–(23) and respects the same commitment-dependent bounds as the base case (24). The post-contingency DC power flow (25) uses the indicator $A_{c,\ell}$, which is zero for the outaged branch and unity otherwise; the corresponding flows are limited by the emergency rating $F_\ell^C$ (26). Nodal balance under every contingency (27) preserves the same time-invariant $P^{\mathrm{HC}}$ power withdrawal at $b^*$. Emergency load shedding is disallowed in *N*-SCUC—a conservative planning-stage choice revisited in Section IV.

Each of the $24 \times 4 = 96$ instances is implemented in Pyomo and solved with Gurobi (MIP gap tolerance 0.5%, time limit 6000 s); all reached optimal or near-optimal solutions within tolerance. Active constraint groups for each formulation are summarized in Table I.

## III. Results and Discussion

### A. Full Bus-Level HC Comparison

Figure 1 presents the complete HC results for all 24 buses. Panel (a) plots bus-level HC, with the 138-kV (Buses 1–10) and 230-kV (Buses 11–24) sub-networks separated by a vertical dashed line; Panel (b) shows the per-bus HC difference between DC-OPF and *N*-SCUC, i.e., the overestimation introduced by the simplified model. The most accurate model, *N*-SCUC, yields the lowest system-wide average HC at 444 MW; relaxing security raises this to 532 MW when *N*-1 contingency constraints are dropped (R-SCUC), to 630 MW when spinning reserve is also removed (B-SCUC), and to 628 MW for DC-OPF. The slight drop from B-SCUC to DC-OPF reflects that DC-OPF keeps all units online, enforcing a minimum-generation floor; otherwise, the average rises steadily as security layers are removed, confirming that *N*-1 contingency and spinning reserve are the dominant limiting factors while UC logic exerts negligible aggregate influence. The reserve effect is structural: constraint (7) enforces a loss-of-largest-unit criterion, requiring the system to hold reserve equal to the output of the largest committed unit (400 MW). This withheld capacity imposes a hard ceiling on HC, evident in the many buses that saturate at 721.73 MW under R-SCUC and *N*-SCUC, exactly the total capacity (3,405 MW) minus this reserve (400 MW) minus the peak load (2,283 MW).

Buses 16 and 18 exhibit the largest absolute overestimation, approximately 332 MW each (highlighted in Fig. 1(b)). This wide spatial variation motivates the bus-level analyses in the following subsections. We caution that absolute MW values, rather than percentages, provide the more meaningful basis for comparison, since several low-HC buses produce extreme percentages from very small baselines.

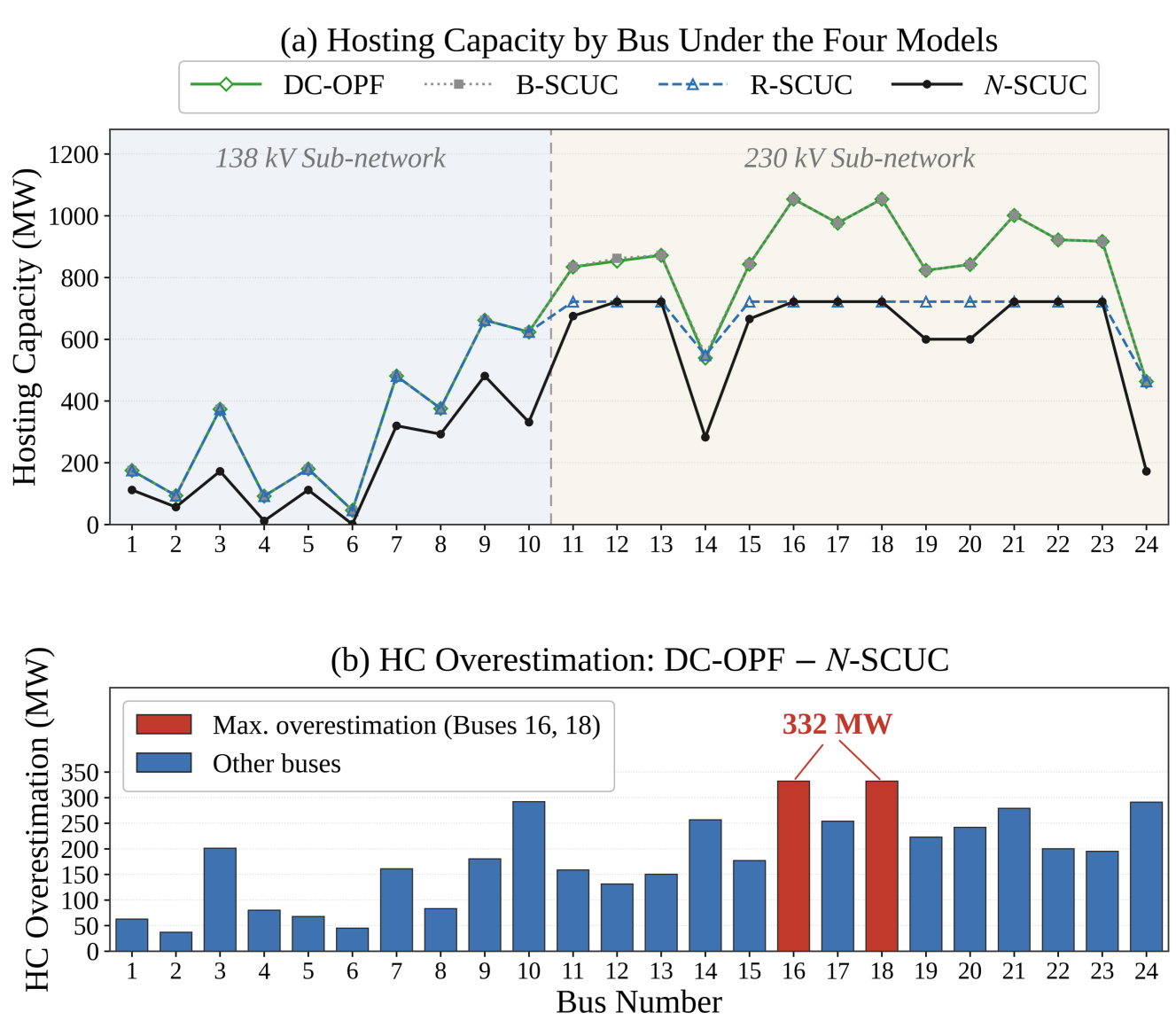

Fig. 1. Hosting capacity results for all 24 buses. (a) HC values under the four models (DC-OPF, B-SCUC, R-SCUC, and *N*-SCUC), ordered by increasing fidelity. Shaded regions indicate the 138 kV (Buses 1–10) and 230 kV (Buses 11–24) sub-networks. (b) Absolute HC overestimation of the simplified DC-OPF relative to the *N*-SCUC for each bus.

### B. Bus-Level Analysis

The HC distribution across the network is highly non-uniform, reflecting the interplay of transmission thermal capacity, network redundancy, and local generation adequacy. Buses in the 230-kV sub-network (Buses 11–24) generally exhibit larger HC than those in the 138-kV sub-network (Buses 1–10). The difference between the two voltage levels enters the model through the line thermal ratings, which the 230-kV sub-network carries at substantially higher values (with normal and emergency thermal limits averaging about 444 and 554 MW, versus 158 and 195 MW for the 138-kV lines), giving greater transfer capability and hence higher HC. Under the accurate *N*-SCUC, eight buses (12, 13, 16–18, 21–23) reach a common HC ceiling of 722 MW, indicating a shared system-level bottleneck. Dropping *N*-1 constraints (R-SCUC) brings four more buses (11, 15, 19, 20) to this ceiling, confirming that post-contingency thermal limits had bound this second group. Under B-SCUC and DC-OPF the ceiling rises sharply, with Buses 16 and 18 reaching 1,053–1,054 MW, Bus 21 reaching 1,001 MW, and Bus 17 reaching 976 MW—the system-wide spinning reserve requirement (7) that constrained these well-connected high-voltage buses no longer being enforced.

To isolate the contribution of each security layer, the three sequential simplification steps are examined relative to the accurate *N*-SCUC. Dropping *N*-1 (*N*-SCUC to R-SCUC) produces the most pronounced relative overestimation, concentrated at weakly-connected buses. Its largest absolute increases are split across both sub-networks—Bus 10 (138 kV) and Bus 24 (230 kV) rise by 292 and 291 MW, followed by Bus 14 (230 kV) at 266 MW; within the 138-kV sub-network the largest effects occur at Buses 10, 3, and 9 (292, 201, and 181 MW). In relative terms the effect is most dramatic at Bus 6 (HC rises from approximately 1 MW to 46 MW) and Bus 4 (12 to 92 MW); both have only two connecting lines and no local generation, making the post-contingency constraints (22)–(27) particularly binding. Dropping spinning reserve (R-SCUC to B-SCUC) produces the largest absolute overestimation of the three steps (system-average 97 MW, versus 88 MW for the *N*-1 step): it predominantly inflates high-voltage buses previously capped at 722 MW—Buses 16 and 18 by 332 MW each and Bus 21 by 280 MW—while lower-voltage buses change negligibly, their HC already set by local thermal constraints (20). The mechanism is twofold: each committed unit must withhold capacity for reserve (constraint (6)), reducing the generation headroom available for the additional load, and the envelope constraint (7) requires total reserve to cover the loss of the largest committed unit and its reserve. The effect concentrates at well-connected high-voltage buses, where transfer capability is otherwise ample and reserve becomes binding. Dropping UC (B-SCUC to DC-OPF) changes most buses little; notably, HC at Buses 12 and 14 is about 9 MW lower under DC-OPF, because DC-OPF cannot decommit units and the resulting aggregate minimum-generation floor constrains dispatch more than selective decommitment under SCUC—a counter-intuitive result showing the HC–fidelity relationship is not strictly monotonic.

Network topology, rather than load magnitude, primarily governs HC. Bus 6 (base load 148 MW) records the lowest HC across all models despite not being the most heavily loaded, owing to its two-line connectivity and lack of local generation. By contrast, Bus 9 carries the highest base load (175 MW) yet reaches 481 MW under *N*-SCUC, supported by five lines (to Buses 3, 4, 8, 11, and 12) totaling 1,192.5 MW of Rate A capacity. Transmission redundancy is thus a more reliable predictor of HC than load level alone.

### C. Line Utilization and Computational Performance

In this steady-state DC analysis, 94 of the 96 instances have at least one line loaded to its normal thermal limit under pre-contingency operation, so a base-case thermal limit (20) is usually what caps HC. However, two patterns show that a fully loaded line is not always the true limiting factor. First, at Buses 19 and 21 under *N*-SCUC, no line reaches its normal limit (the highest loadings are 98.1% and 99.0%); here HC is set instead by the post-contingency limits (26) and the system-wide reserve requirement (7). Second, even when a line is fully loaded in the base case, it need not be the only binding constraint: at low-HC buses such as Bus 6, the post-contingency limits (26) bind at the same time and are the main reason HC is so low.

Figure 2 summarizes the computational performance. Panel (a) shows the average per-bus solve time on a logarithmic scale with total wall-clock time overlaid, while Panel (b) presents the average and maximum MIP optimality gaps against the 0.5% solver tolerance. All instances were solved on a workstation with an Intel Core i7-12700 processor (12 cores, 20 threads) and 32 GB of RAM, using Gurobi 12.0.3 under Python 3.12 with Pyomo (6.10).

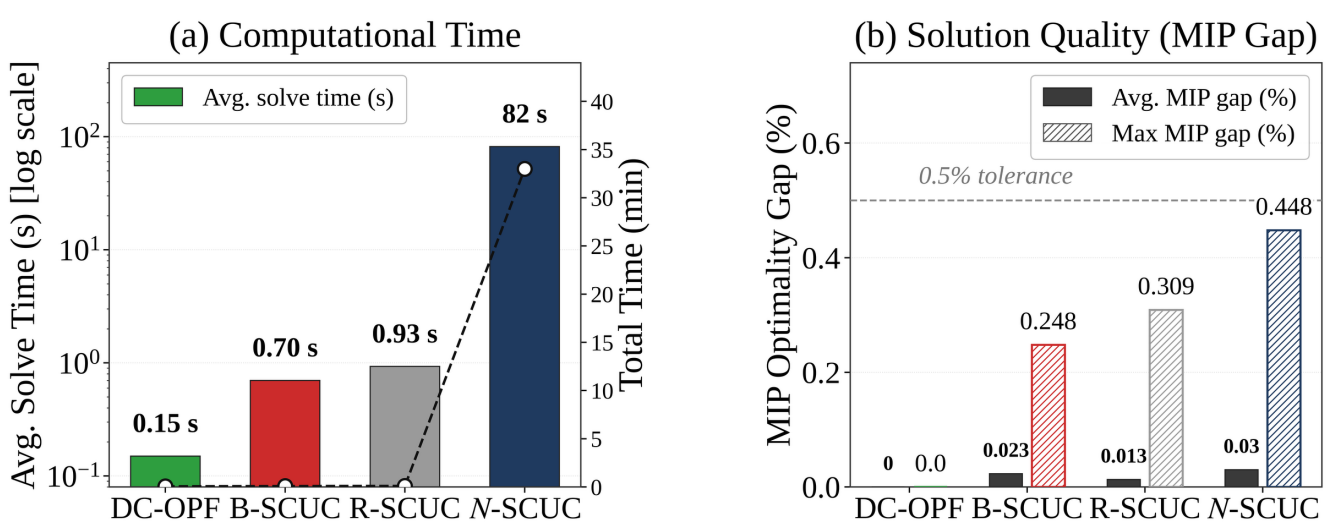

Fig. 2. Computational performance and Solution Quality of the four models across all 24 buses. (a) Average per-bus solve time (log scale) and total time. (b) Average and maximum MIP optimality gaps; the dotted line marks the 0.5% solver tolerance.

The computational cost rises steeply with model fidelity (Table I). The *N*-1 sub-block dominates: its 37 contingency scenarios account for 93.9% of the variables and 95.6% of the constraints of the *N*-SCUC, which averages 81.8 s per bus (32.7 min total across the 24 buses). Omitting *N*-1 (*N*-SCUC to R-SCUC) removes this sub-block, cutting the solve time to 0.93 s but discarding the post-contingency security that makes *N*-SCUC accurate. Further dropping spinning reserve (R-SCUC to B-SCUC) yields only a 14.5% model-size reduction and a modest drop to 0.70 s, while removing the 2,376 binary commitment variables (B-SCUC to DC-OPF) converts the problem into a linear program solved in 0.15 s.

## IV. Conclusion

This paper quantifies the effect of progressive security-constraint relaxation on hosting capacity across all 24 buses of the modified IEEE RTS system, under a planner-oriented definition in which a single time-invariant load is added at one bus at a time. Two security layers dominate the hosting limit, with distinct spatial signatures: *N*-1 branch-contingency enforcement governs the 138-kV sub-network, whereas spinning reserve governs the well-connected 230-kV sub-network; unit commitment has a negligible aggregate effect, and the per-bus magnitudes are reported in Section III. The DC-OPF and B-SCUC substantially overestimate HC, so their low computational cost does not make them useful substitutes for the *N*-SCUC. Where the *N*-SCUC is intractable and a simplified model must be used, the simplified-model HC should be treated as an optimistic upper bound and corrected downward for the omitted *N*-1 security.

The most immediate extension is the addition of an AC power-flow stage to evaluate bus voltage magnitudes and reactive-power limits, capturing the voltage-related constraints the DC model omits. Beyond steady-state assessment, the framework can co-optimize hosting capacity with transmission and generation upgrade planning to identify the most cost-effective reinforcements at the binding buses. Further extensions include interconnection-scale systems with renewable variability, generator *N*-1 contingencies, and bounded emergency load shedding.

## Acknowledgment

This research was supported in part by the University of Houston Energy Transition Institute Seed Grant Program.